\documentclass[twoside,10pt]{article}
\usepackage{1}
\title{\Large \bf Status of the NICA/MPD project}
\author{Kolesnikov V.I.$^1$ and Zinchenko A.I.$^1$ for the MPD Collaboration}
\date{}
\begin{document}
\maketitle

\begin{center}
\vspace*{-0.3cm}
{\it  $^1$~Joint Institute for Nuclear Research (JINR)}
\end{center}

\vspace{0.3cm}

\begin{center}
{\bf Abstract}\\
\medskip
\parbox[t]{10cm}{\footnotesize
A general-purpose detector for studying heavy-ion collisions
at the NICA facility is under construction at JINR. The NICA/MPD
physics program, basic design requirements, and the MPD experimental
setup will be described. Results of detector simulation and the expected
performance for selected observables will be presented.}
\end{center}

\section{Introduction}
\vspace{-2mm}
\hspace{5mm}Experimental studies of QCD matter at high baryon densities
provides new perspectives to resolve the most fundamental problems of the
underlying theory - confinement and chiral symmetry breaking.
Our knowledge about the QCD phase structure at intermediate $\mu_B$ is
poor: theory suggests a first-order transition at large $\mu_B$ and its turn
into a crossover at small baryon densities (and high $T$), hence, for consistency,
the critical endpoint (CEP) is expected to exist. However a rigorous proof on
such a QCD structure is not yet available and new reliable
experimental data on the nature and properties of the phase transition are
needed.

The goal of the NICA research program at JINR is to investigate 
a wide range of physics phenomena in heavy-ion collisions including phases
of nuclear matter and EoS at high baryon density, properties of the hadron
spectral function and features of hyperon-nucleon interaction in the medium,
critical behavior of the QCD matter  and the spin structure of the
nucleon~\cite{nica}.    
The new NICA facility\cite{nica_cdr} will be capable to provide ion beams with
the design luminosity of $10^{27}$~cm$^{-2}$c$^{-1}$ (for gold ions) in the energy
range  from $\sqrt{s}=4$ to 11A~GeV. In 2018, we will start a detailed
energy and system size scan focusing on hadroproduction and dilepton studies,
event-by-event fluctuations and correlations.
Production of composite objects with strangeness (hypernuclei) are of particular
interest, since they are a unique tool to probe new nuclear structures or unknown
properties of the baryonic interaction, which cannot be seen from the study of
ordinary nuclei.
 \begin{figure}[t]
\begin{center}
\includegraphics[width=6cm]{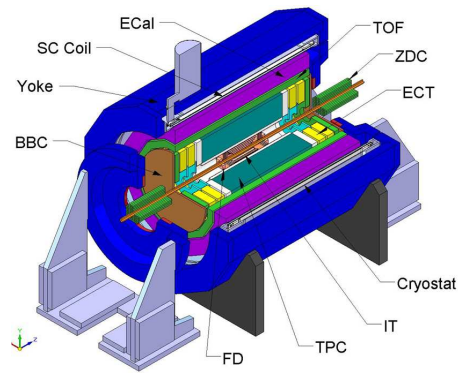}
\end{center}
\vspace{-0.5cm} \caption{Schematic view of the MPD detector.} \vspace{-0.2cm}
\label{fig_mpd}
\end{figure}
\section{MPD detector}
\vspace{-2mm}
\hspace{5mm}The {\bf M}ulti{\bf P}urpose {\bf D}etector (MPD) is designed to fully
exploit the NICA physics potential. It is a spectrometer with a large
uniform acceptance (full azimuth) capable of detecting and identifying hadrons,
electrons and gammas  at the very high event rate achieved at NICA~\cite{mpd_nim}.
All the elements of the detector (see Fig.\ref{fig_mpd}) are ordered inside a
superconducting solenoid generating a magnetic field of up to 0.6 T.
Tracking will be performed with a cylindrical Time Projection Chamber (TPC)
with a MWPC-based readout. The TPC is required to have a high efficiency
and momentum resolution over the pseudorapidity range $|\eta|<2$.
Having of about 65 measured space points per a track, TPC will enable particle
identification via the specific energy loss ({\it dE/dx}) measurement  with a
precision better than 8\%.
At large pseudorapidities TPC tracking will be supplemented by a multi-layer straw
tube tracker (ECT) located just after the TPC end plates.
The Inner Tracker (IT) will consist of four layers of double-sided silicon microstrip
detectors serving mainly for determination of the position of the primary
interaction vertex and secondary decay vertices.   
The Time-Of-Flight (TOF) system made by RPC (Resistive Plate Chambers) is intended
for charged hadron identification. The TOF detector covers
$|\eta|<3$ and its performance should allow the separation of kaons from
protons up to a total momentum of 3 GeV/c. Behind the TOF detector, a high
segmented electromagnetic calorimeter (ECAL) for electron and gamma
identification  will be located.  Arrays of quartz counters (FD) are meant 
for fast timing and triggering, and two sets of hadron calorimeters (ZDC),
covering the pseudorapidity region $2.5<|\eta|<4$,  will measure the
forward going energy for centrality selection and event plane analysis.
A more detailed description of the detector components can be found
elsewhere~\cite{mpd_cdr}.
\section{MPD performance studies and R\&D}
\begin{figure}
\begin{center}
\begin{minipage}{10pc}
\includegraphics[width=10.5pc]{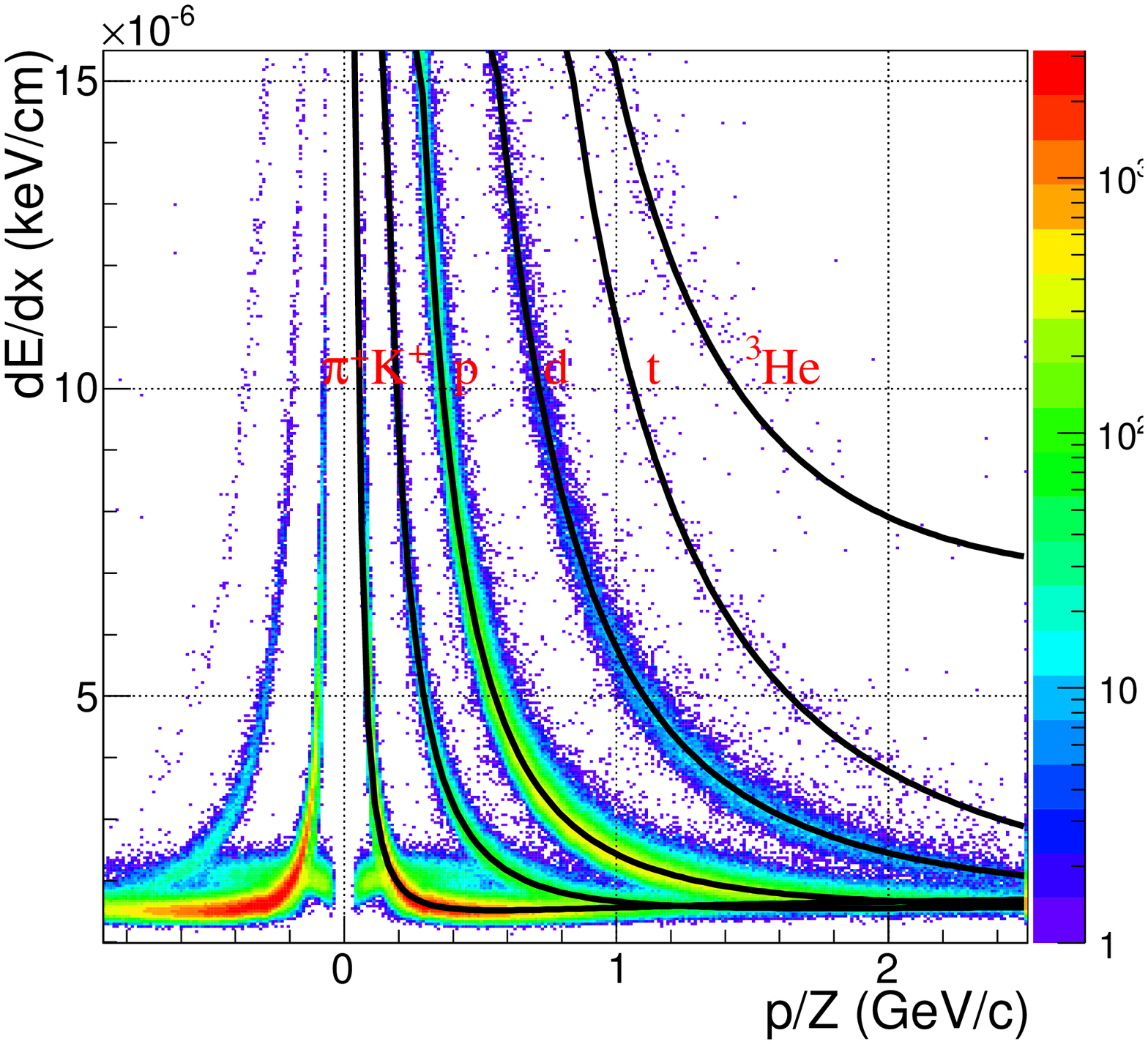}
\end{minipage}\hspace{0.5pc}
\begin{minipage}{13pc}
\includegraphics[width=13.5pc]{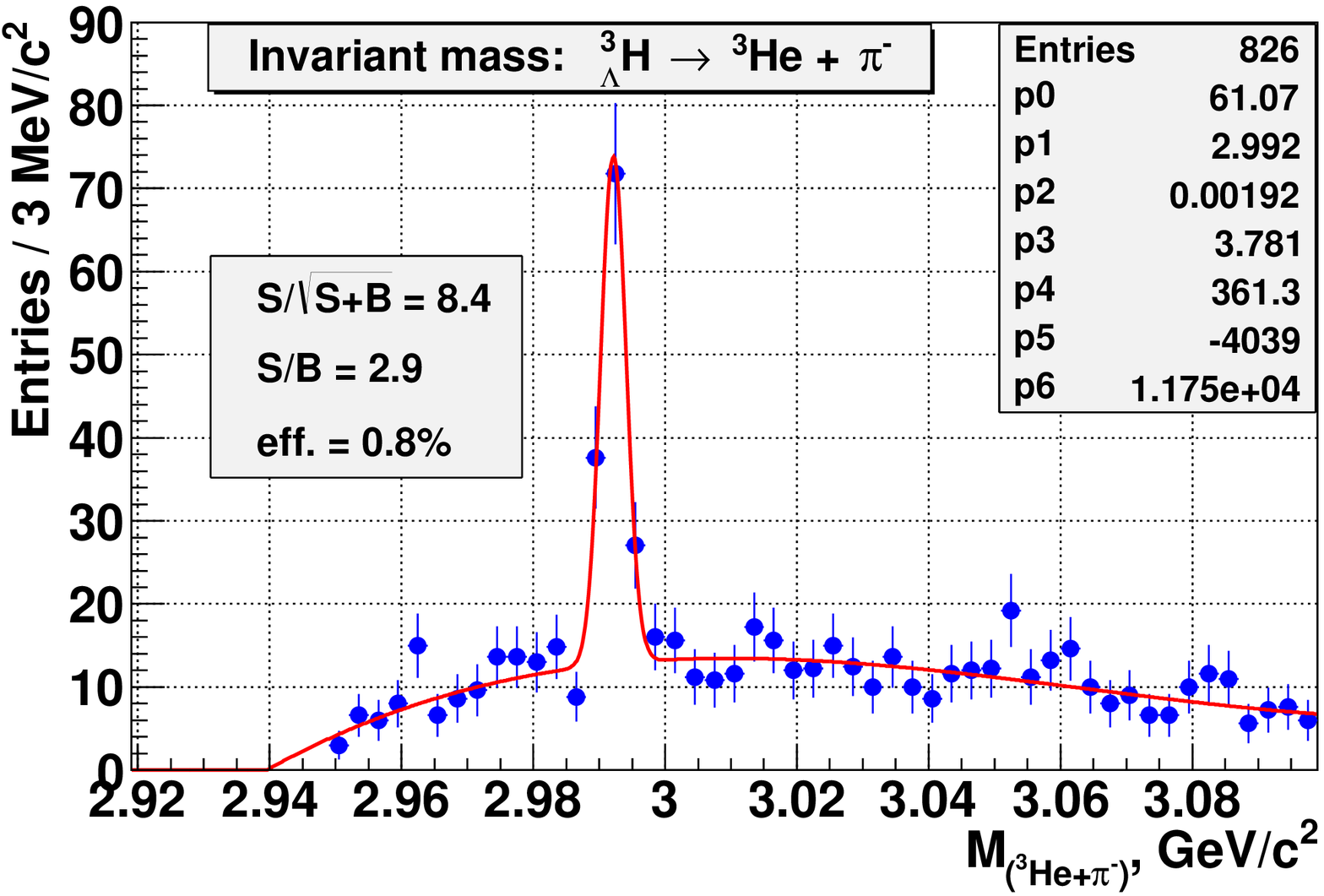}
\end{minipage}
\caption{\label{fig_hyper}(Left) Specific energy loss of $\pi$, K, p, d, t
 and $^3He$ in the MPD TPC gas mixture. PID selection is based on a 3$\sigma$
band relative to the $dE/dx$-paramemeterization shown by the solid line.
(Right) Invariant mass distribution
 of $^3He$ and $\pi^-$ candidates from central Au+Au collisions at $\sqrt{s}=5$A GeV.
 A Gaussian plus polynomial fit (solid line) is superimposed on the signal
 distribution (symbols)}  
\end{center}
\end{figure}

\vspace{-7pt}
\paragraph{MPD tracking and PID performance.}
\begin{figure}
\begin{center}
\begin{minipage}{13pc}
\includegraphics[width=13pc]{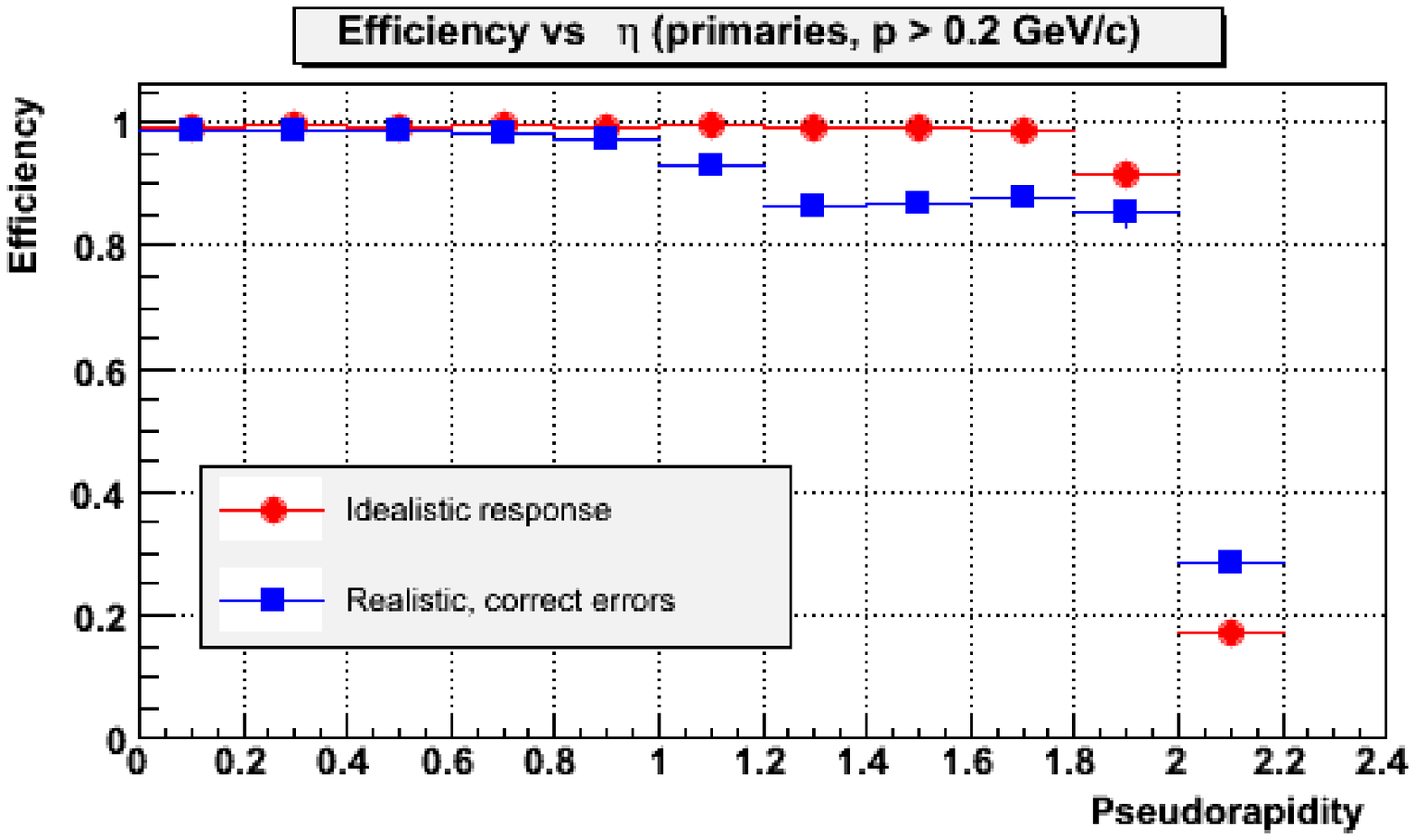}
\end{minipage}
\begin{minipage}{11pc}
\includegraphics[width=11pc]{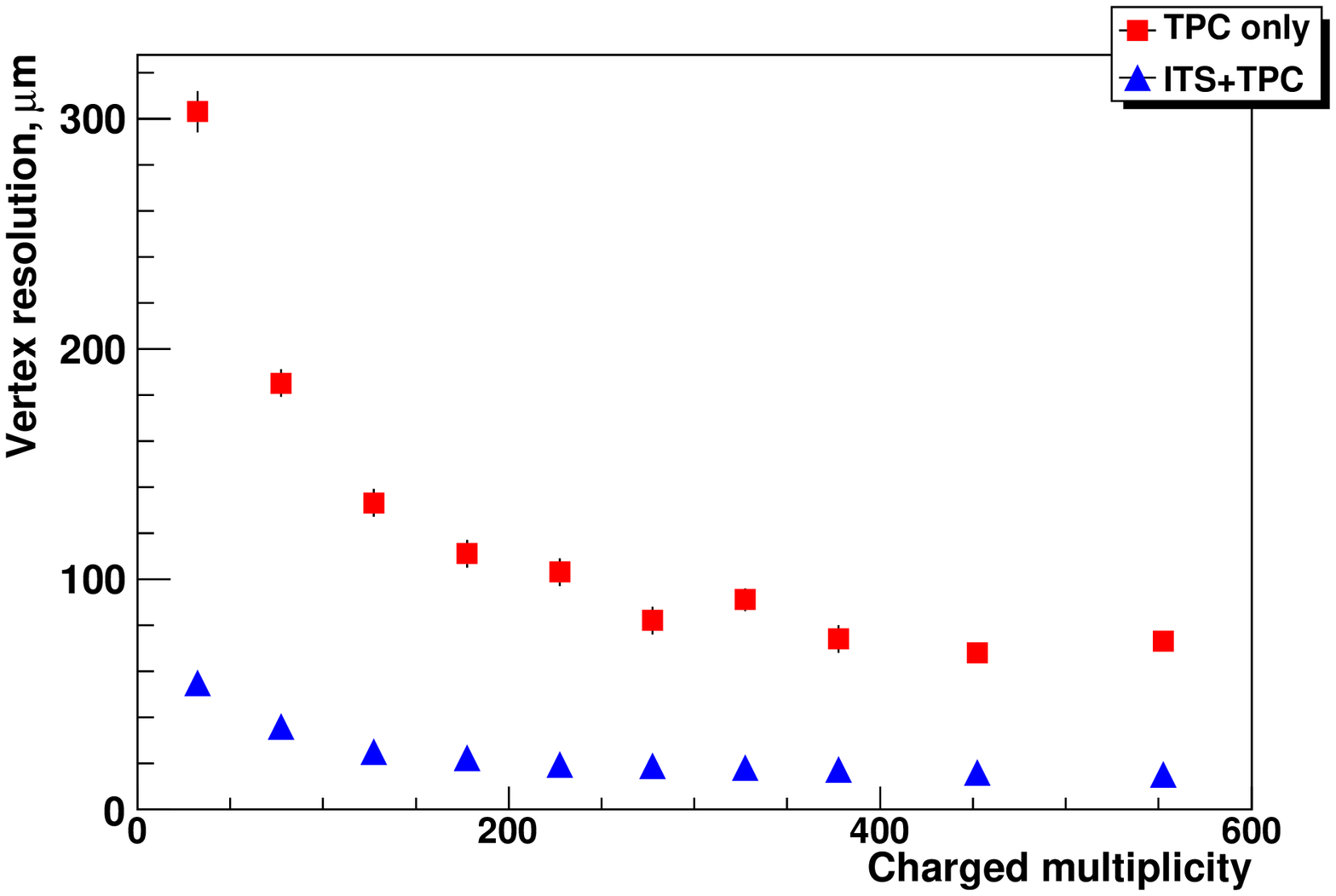}
\end{minipage}\hspace{1pc}%
\caption{\label{fig_eff} (Left panel) Tracking efficiency in TPC versus $\eta$
for an ideal and realistic TPC response. (Right) Primary vertex reconstruction
resolution ($\sigma_z$) as a function of charged track multiplicity.}  
\end{center}
\end{figure}
The MPD performance studies were performed within the MPDRoot framework ~\cite{mpd_cdr},
which provides an interface to external event generators (like UrQMD), transport codes
(Geant3,4), and implements MPD detector response simulations and event reconstruction
algorithms.
Tracking and vertexing performance of MPD are shown in Fig.\ref{fig_eff}
for single track efficiency and spatial resolution along the beam
axis $\sigma_z$. 
As one can see, the detector is able to provide highly efficient tracking up to
$\eta=2$, and a resolution better than 40 microns can be achieved
in central collisions.
\begin{figure}
\begin{center}
\begin{minipage}{10pc}
\begin{minipage}{10pc}
\includegraphics[width=10pc]{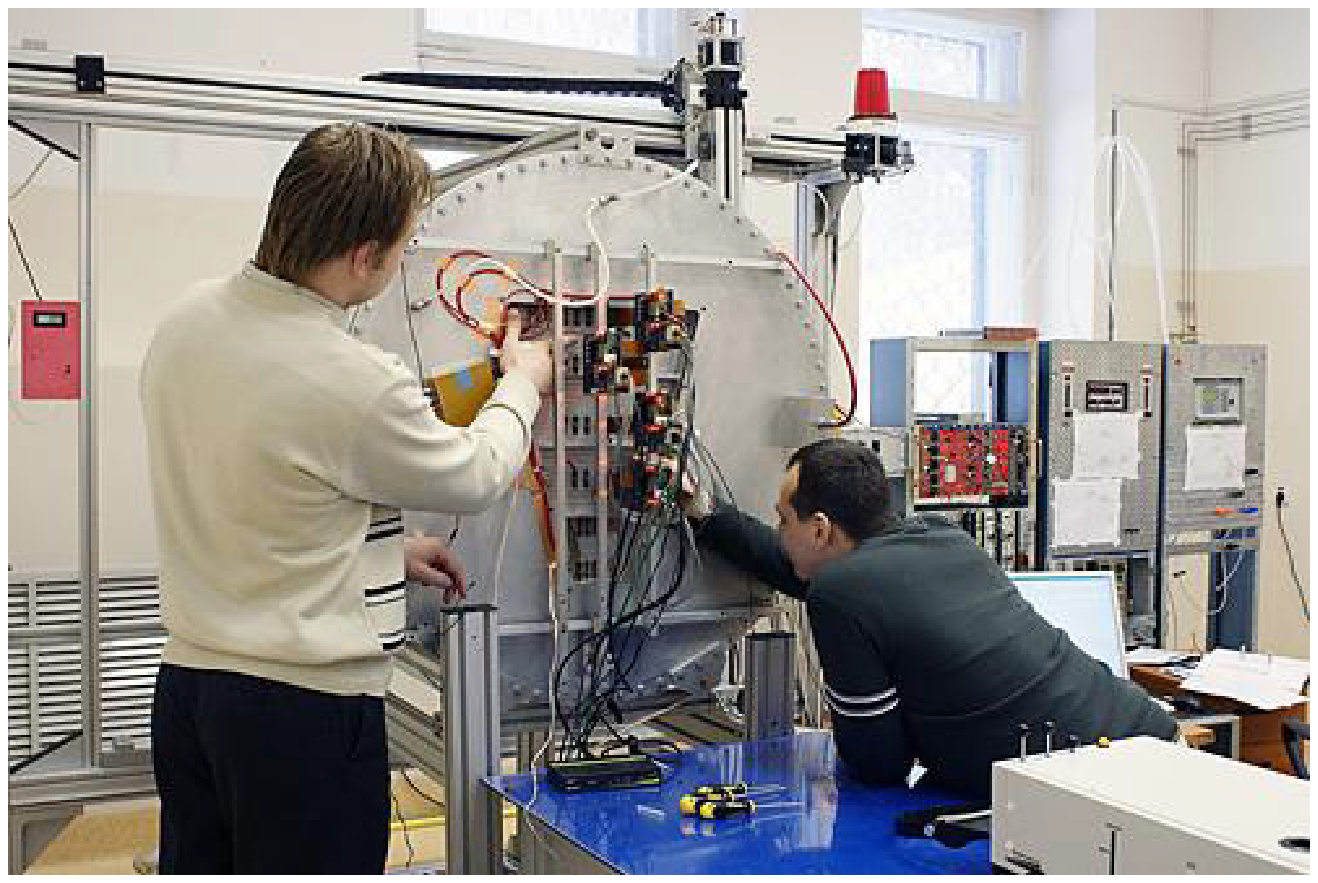}
\end{minipage}
\begin{minipage}{10pc}
\includegraphics[width=10pc]{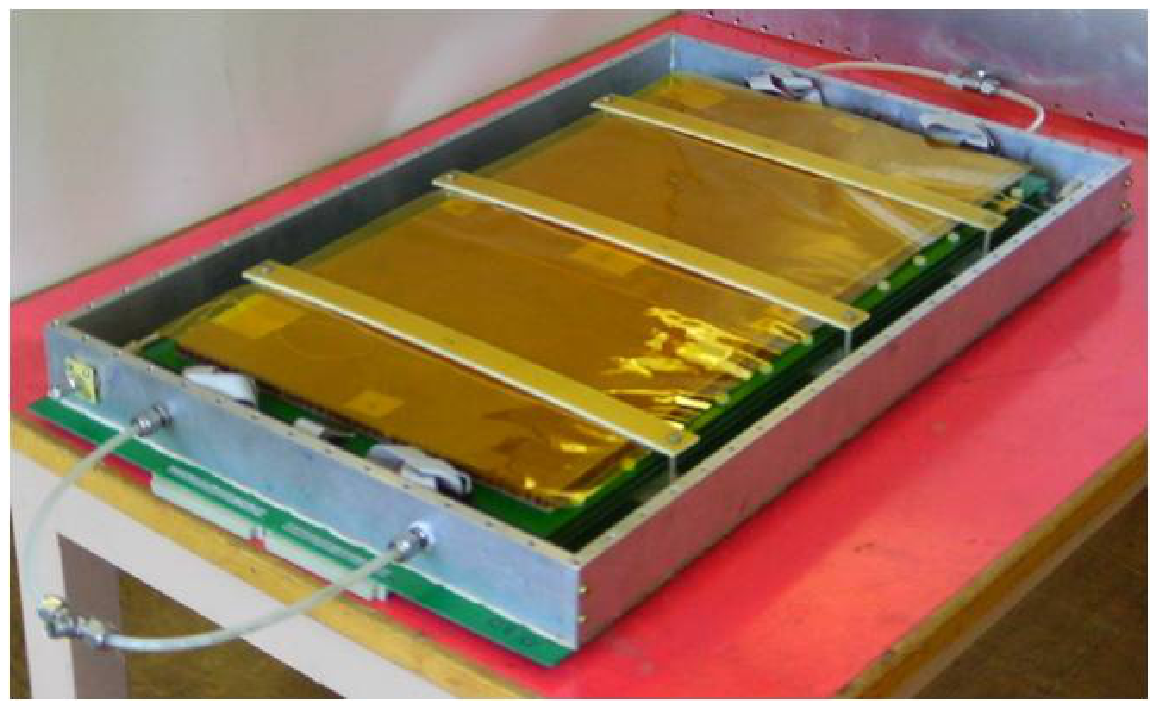}
\end{minipage}
\end{minipage}\hspace{2pc}%
\begin{minipage}{10pc}
\includegraphics[width=10pc]{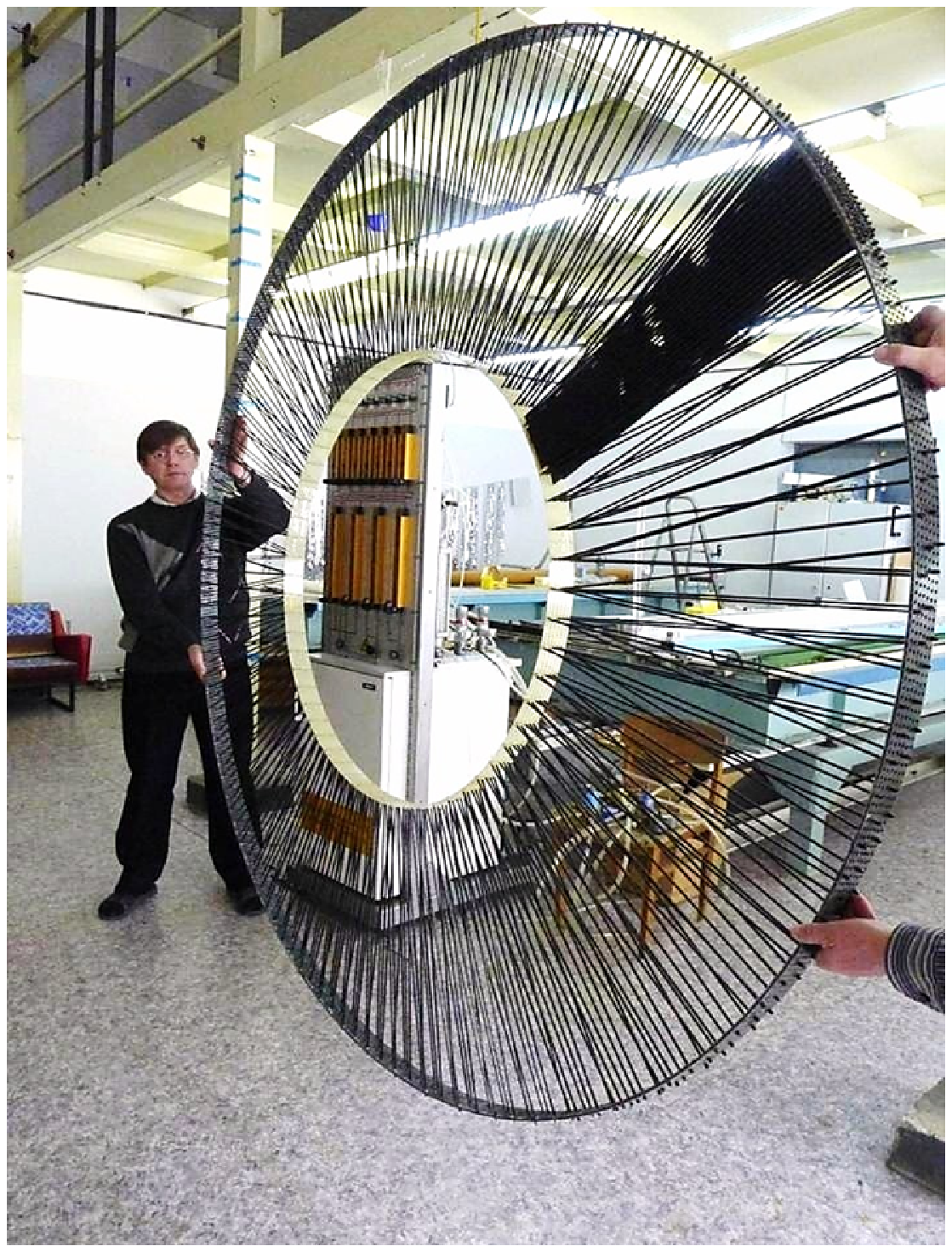}
\end{minipage}\hspace{1pc}%
\caption{\label{fig_proto} (Upper left) A first assembled TPC prototype in
laboratory measurements. (Bottom left) A full-scale TOF module. (Right)
One of the ECT endcap wings.}  
\end{center}
\end{figure}
\vspace{-10pt}
\paragraph{Progress in MPD prototyping}
During the years 2012-13, the main MPD R\&D activities were aimed at developing of
novel techniques in construction of TPC, TOF, IT and ECT detectors, as well as at
production and tests of the first detector prototypes (see Fig.~\ref{fig_proto}).
For example, to ensure lightweight and mechanical stability of the TPC, its
supporting elements will be made from composite materials in collaboration with
industry. Also, to ensure accurate tracking at large-$\eta$, a new construction
technology for ECT modules was developed allowing the alignment of straw tubes
with a 100 micron precision within an object of 2 meters in diameter.    
\vspace{-10pt}
\paragraph{MPD potential for hypernuclei measurements.}
The feasibility of precise hypernuclei measurements at NICA has been investigated
with the event generator DCM. The model implements a coalescence-based algorithm
for (hyper)nuclei formation and calculated yields of fragments are in a good
agreement with experimental data~\cite{qgsm_2}. Roughly $5\cdot 10^5$
central Au+Au collisions at $\sqrt{s}=5$A GeV were analyzed including full event reconstruction, particle identification by means of combined dE/dx
(Fig.~\ref{fig_hyper}) and TOF measurements, and search for secondary vertices.
To improve the signal-to-background ratio, a set of quality and topological cuts
were applied on the number of TPC space points and the distance between the daugthers
at the decay vertex.
The results for $^3_{\Lambda}He \rightarrow ^3He + \pi^-$ are shown in
Fig.\ref{fig_hyper} (right panel).
With the overall reconstruction efficiency of about 1\% and design NICA luminosity we
expect roughly 500 reconstructed $^3_{\Lambda}He$ candidates per day of data
taking. Such high event rates provide a good opportunity to gain further insights into production mechanism and properties of hypernuclei.

\end{document}